\documentclass[reprint,twocolumn,superscriptaddress,amsmath,amssymb,aps,prl]{revtex4-1}

\usepackage{graphicx}
\usepackage{graphics}
\usepackage{amsmath}
\usepackage{dcolumn}
\usepackage{bm}

\begin{document}


\title{Stable structures and electronic properties of perovskite oxide monolayers}

\author{Xiang-Bo Xiao}
\affiliation{Beijing National Laboratory for Condensed Matter Physics, Institute of Physics, Chinese Academy of Sciences, Beijing 100190, China}
\affiliation{School of Physical Sciences, University of Chinese Academy of Sciences, Beijing 100190, China}
\author{Bang-Gui Liu}\email{bgliu@iphy.ac.cn}
\affiliation{Beijing National Laboratory for Condensed Matter Physics, Institute of Physics, Chinese Academy of Sciences, Beijing 100190, China}
\affiliation{School of Physical Sciences, University of Chinese Academy of Sciences, Beijing 100190, China}

\date{\today}

\begin{abstract}
It is highly desirable to search for promising two-dimensional (2D) monolayer materials for deep insight of 2D materials and applications. We use first-principles method to investigate tetragonal perovskite oxide monolayers as 2D materials. We find four stable 2D monolayer materials from SrTiO$_3$, LaAlO$_3$, KTaO$_3$, and BaFeO$_3$, denoting them as STO-ML, LAO-ML, KTO-ML, and BFO-ML. Our further study shows that through overcoming dangling bonds the first three monolayers are 2D wide-gap semiconducotors, and BFO-ML is a 2D isotropic Heisenberg ferromagnetic metal. There is a large electrostatic potential energy difference between the two sides, reflecting a large out-of-plane dipole, in each of the monolayers. These make a series of 2D monolayer materials, and should be useful in novel electronic devices considering emerging  phenomena in perovskite oxide heterostructures.
\end{abstract}

\pacs{Valid PACS appear here}
\maketitle



The advent of graphene stimulates huge interest in two-dimensional (2D) materials\cite{graphene,graphene1}.
Besides graphene, many free-standing 2D monolayer materials have been found, such as hexagonal BN, transition-metal dichalcogenides, metal halides, and so on\cite{hbn,mos2,rev1,rev2}. As a key feature, these monolayer materials have corresponding layered three-dimensional (3D) bulk materials and can be made by exfoliation from the layered 3D materials. They can host many interesting electronic, optical, magnetic, mechanical, and topological properties, and can be applied for atomic-thin transistors, spintronic and optical devices, and various novel devices\cite{rev1,rev2,hall,mos2a,toprev}. Furthermore, it was shown by first-principles investigation that there exist a stable double-layer honeycomb structure for majority of traditional binary semiconductors and this honeycomb structure can host interesting topological properties in some cases\cite{zhangsb}. On the other hand, atomically-thin 2D electron/hole gases can be formed in interfaces and surfaces of perovskite oxide heterostructures, where perovskite oxide layers can exist with thickness down to a few unit cells\cite{perov,perov1}. Controllable 2D electronic systems and emerging magnetism, ferroelectricity, and superconductivity can appear in such heterostructures made from otherwise insulating materials such as SrTiO$_3$ and LaAlO$_3$\cite{perova,perova1,perova2}. Surprisingly, free-standing ultrathin (001) films of SrTiO$_3$ and BiFeO$_3$ with thickness down to one unit cell are experimentally realized, which implies that tetragonal perovskite oxide monolayers (one-unit-cell-thick (001) layers) may be structurally stable\cite{perov2d}.

Here, we construct tetragonal perovskite oxide monolayers as 2D materials, and investigate their structural stability and electronic structures. As a result, we find that SrTiO$_3$, LaAlO$_3$, KTaO$_3$, and BaFeO$_3$ monolayers are structurally stable, and denote them by STO-ML, LAO-ML, KTO-ML, and BFO-ML, respectively. Considering possible dangling bonds on the surfaces, it is surprising that STO-ML, LAO-ML, and KTO-ML are 2D wide-gap semiconductors. Actually, substantial atomic relaxations remove possible dangling bonds in STO-ML, LAO-ML, and KTO-ML. For BFO-ML, we obtain a 2D ferromagnetic metal due to dangling bonds.
For each of the four monolayers, we find a large electrostatic potential energy difference between the two sides, which means a large dipole moment perpendicular to the monolayer plane. Some of the bond lengths are substantially different from those of the corresponding bulk values. More detailed results will be presented in the following.


Our calculations are done with the projector augmented wave (PAW) \cite{paw} method within the density functional theory \cite{dft1,dft2}, as implemented in the vienna ab initio simulation package (VASP) \cite{vasp1,vasp2}. We construct one-unit-cell (001) layers, or monolayers, of four perovskite oxides (ABO$_3$), namely SrTiO$_3$, LaAlO$_3$, KTaO$_3$ and BaFeO$_3$, and denote them by STO-ML, LAO-ML, KTO-ML and BFO-ML, respectively. For each of the cases, we adopt similar computational slab model consisting of the monolayer and a vacuum layer with the thickness more than 18 \AA. The $z$ axis is along the [001] direction, perpendicular to each of the monolayers. In each monolayer, the upper atom layer (AL) consists of A and O$_1$, and the lower AL is amde by B and O$_2\times 2$. We take the generalized-gradient approximation (GGA) of Perdew-Burke-Ernzerhof (PBE) \cite{pbe} for the exchange-correlation potential. We consider the electron correlation of Fe atoms by using GGA+U method, with $U = 5.0$ eV and $J = 2.0$ eV \cite{stc1}. For the $\Gamma$-centered grids of k-points, we use $8\times8\times1$ to optimize the crystal structure of the slab model and $10\times10\times1$ to calculate the total energies. The plane wave energy cutoff is set to 600 eV. Our convergence standard requires that the Hellmann-Feynmann force on each atom is less than 0.001 eV/\AA{} and the absolute total energy difference between two successive loops is smaller than 10$^{-6}$eV. The spin-orbit coupling is also taken into account to investigate how it effects the electronic properties. In order to compare the electronic properties of these systems with experiment, hybrid functional HSE06 functional is also used for calculations. The phonon dispersion spectra are calculated using the density-functional perturbation theory (DFPT) implemented in Phonopy package \cite{phonon}.


After full structural optimization, the monolayer consists of two atom layers (ALs) at different $z$ coordinates. We show the $z$-dependent average electrostatic potentials of the four slabs in Fig. 1. In each cases, the two minima corresponds to the two ALs in the monolayer. The B-O2 AL is located near $z=0$ and the A-O AL at the upper minimum, as shown in the inserts of Fig. 1. The stability of these monolayers is confirmed by the phonon dispersion curves shown in Fig. 2, because it is clear that their branches of phonon dispersion curves have positive frequencies only. The electrostatic potential energy difference between the two sides should reflect the different work functions ${\Delta\phi}$ between the two ALs in the monolayer. Actually, ${\Delta\phi}$ is proportional to the dipole moment $\mu$ in the $z$ axis of the monolayer. In addition, the electrostatic potential difference implies that there exists an intrinsic built-in electric field across the monolayer along the $z$ direction.

\begin{figure}[!htbp]
{\centering  
\includegraphics[clip, width=8cm]{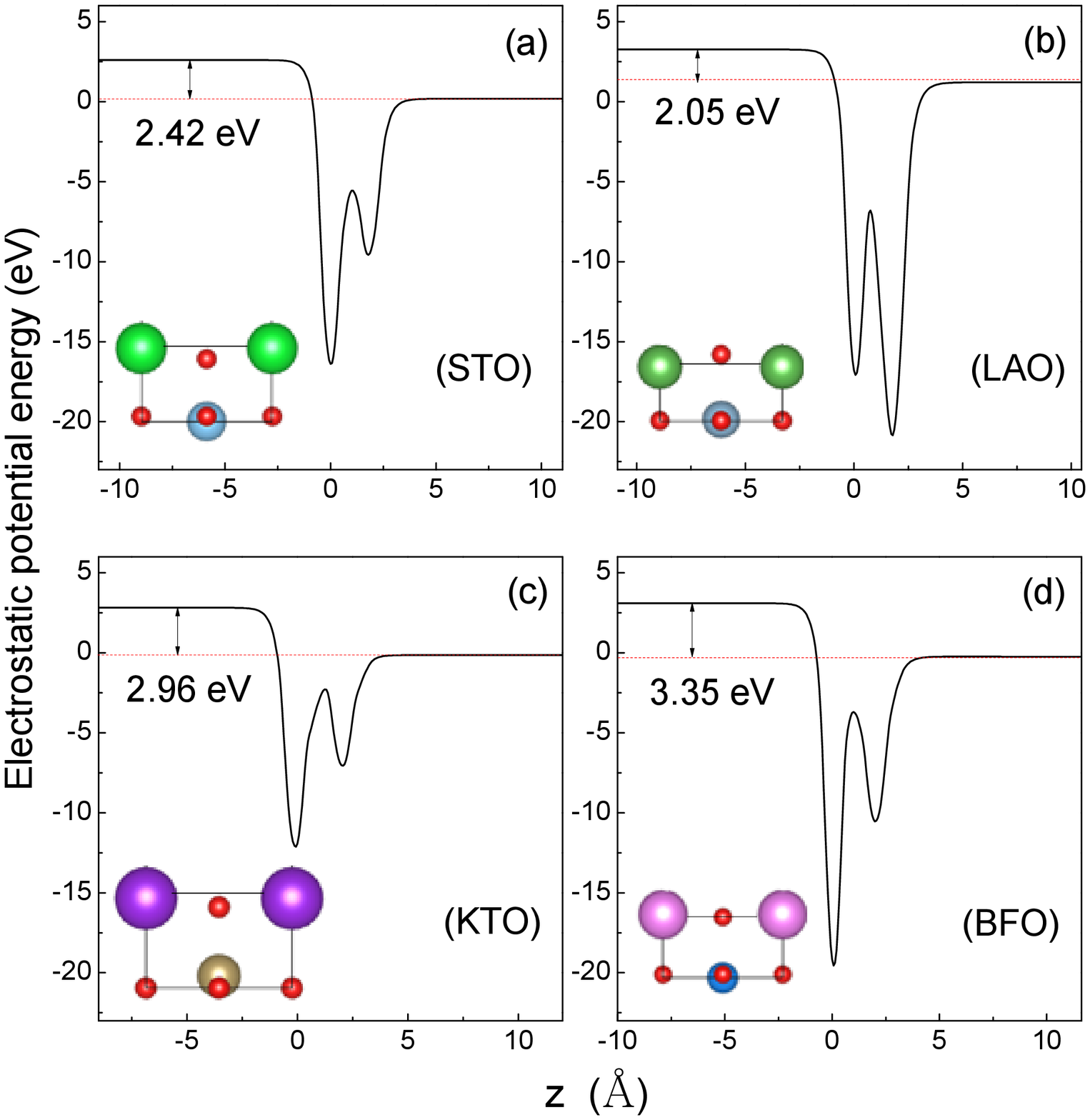}}
\caption{(Color online) The intrinsic average electrostatic potentials of the four perovskite oxide monolayers: STO-ML (a), LAO-ML (b), KTO-ML (c), and BFO-ML (d). The insert describes the atomic structure of the monolayer, with the B-O2 atomic layer corresponding to the minimum at $z=0$.}\label{fig1}
\end{figure}

Presented in Table I are three key parameters for the monolayers: the horizontal lattice constants ($a$), the semiconductor gaps ($G$), and the electrostatic potential energy differences between the two sides ($\Delta\phi$). It is can be seen that for STO-ML, LAO-ML, KTO-ML, and BFO-ML, the theoretical $a$ values are less than the experimental ones $a_e$ by $0.8$\%, 2.3\%, 2.3\%, and 3.5\%, respectively. The first three monolayers (STO-ML, LAO-ML, and KTO-ML) are semiconductive, having semiconductor gaps 1.77 eV, 2.78 eV, and 2.13 eV, respectively. To obtain better estimated values for the gaps, we also calculate the electronic structures by using the hybrid functional HSE06. The structural parameters change a little, but the HSE06 leads to larger gap values: 3.13, 3.89, and 3.46 eV. Our estimation indicates that the true gaps values should remain between the PBE values and the HSE06 ones, but are near the HSE06 values. In contrast, BFO-ML is a ferromagnetic metal with total magnetic moment 4.4 $\mu_B$ per formula unit. As for the electrostatic potential energies, $\Delta\phi$ ranges from 2.05 eV for LAO-ML to 3.35 eV for BFO-ML.

\begin{figure}[!htbp]
{\centering  
\includegraphics[clip, width=8cm]{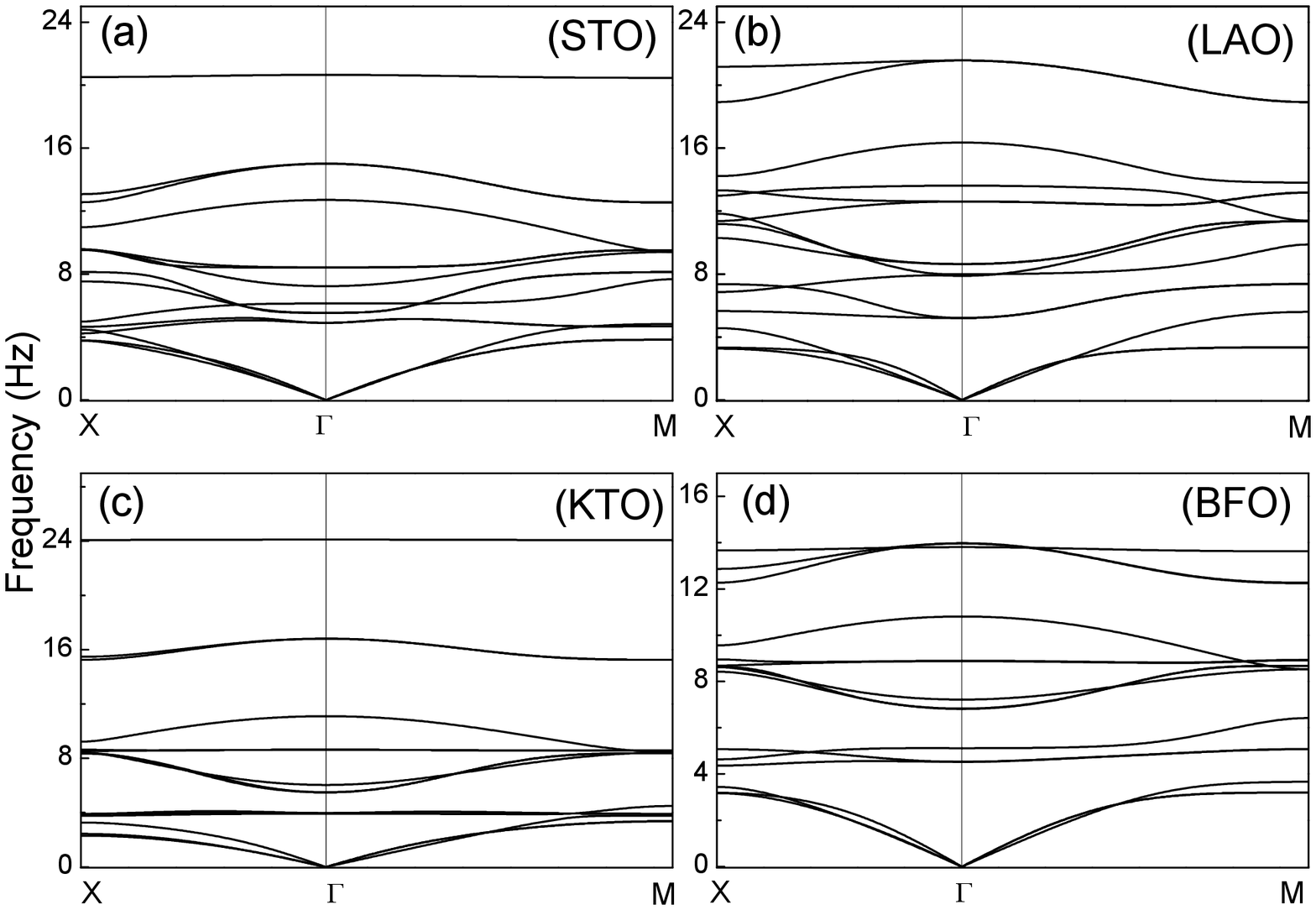}}
\caption{(Color online) The phonon dispersion spectra of the four monolayers:  STO-ML (a), LAO-ML (b), KTO-ML (c), and BFO-ML (d).}\label{fig2}
\end{figure}

\begin{table}[!h]
\caption{The optimized horizontal lattice constants ($a$), the experimental cubic lattice constants ($a_e$), the gaps calculated with PBE (with HSE06) ($G$), and the electrostatic potential energy differences ($\Delta\phi$) of the four monolayers.}
\begin{ruledtabular}
\begin{tabular}{cccccc}
System      & $a$ (\AA)& $a_e$ (\AA) &$G$ (eV)  &  $\Delta\phi$ (eV)  \\ \hline
STO-ML   &  3.875  &  3.905      &  1.77 (3.13)      &  2.42    \\
LAO-ML   &  3.702  &  3.791      &  2.78 (3.89)      &  2.05    \\
KTO-ML   &  3.898  &  3.989      &  2.13 (3.46)      &  2.96    \\
BFO-ML   &  3.831  &  3.971      &  -                &  3.35
\end{tabular}
\end{ruledtabular}
\end{table}

The four bond lengths of the monolayers are summarized in Table II, with the bulk values ($l_{A}$, $l_{B}$) presented for comparison. For STO-ML, the horizontal A-O1 and B-O2 bond lengths ($l_{Ah}$ and $l_{Bh}$) are equal to or little smaller than the corresponding bulk values, the skew A-O2 bond length ($l_{As}$) is larger than $l_{A}$ by 1\%, and the vertical B-O1 bond length ($l_{Bv}$) is 5.6\% smaller than $l_{B}$. For LAO-ML, $l_{Ah}$ and $l_{As}$ are smaller than $l_{A}$ by 5.7\% and 13\%, and $l_{Bh}$ and $l_{Bv}$ deviate from $l_{B}$ by -2.6\% and 3.7\%, respectively. For KTO-ML, $l_{Ah}/l_{A}$ and $l_{As}/l_{A}$ are 0.982 and 1.092, and $l_{Bh}/l_{B}$ and $l_{Bv}/l_{B}$ euqal 0.995 and 0.925, respectively, which implies that $l_{As}$ expands by 9.2\% and $l_{Bv}$ shrinks by 7.5\%. For BFO-ML, the four bond lengths are a little smaller than the corresponding bulk values. These can be seen in the inserts in Fig. 1.

\begin{table}[!h]
\caption{The calculated values of horizontal A-O1 bond lengths ($l_{Ah}$) (A=Sr, La, K and Ba), skew A-O2 bond lengths ($l_{As}$), horizontal B-O2 bond lengths ($l_{Bh}$) (B=Al, Fe, Ti and Ta), and vertical B-O1 bond lengths ($l_{Bv}$) in the four monolayers. The corresponding experimental values in the bulk phases are denoted by $l_A$ and $l_B$.}
\begin{ruledtabular}
\begin{tabular}{ccccccc}
System      & $l_{Ah}$ (\AA) & $l_{As}$ (\AA) & $l_{A}$ (\AA) & $l_{Bh}$ (\AA) & $l_{Bv}$ (\AA) & $l_{B}$ (\AA) \\ \hline
STO-ML   &  2.76          &  2.79   &  2.76      &  1.94       &  1.84    &  1.95      \\
LAO-ML   &  2.66          &  2.46   &  2.68      &  1.85       &  1.97    &  1.90      \\
KTO-ML   &  2.77          &  3.08   &  2.82      &  1.98       &  1.84   &  1.99       \\
BFO-ML   &  2.71          &  2.71   &  2.81      &  1.92       &  1.97   &  1.98
\end{tabular}
\end{ruledtabular}
\end{table}

\begin{figure}[!htbp]
{\centering  
\includegraphics[clip, width=8cm]{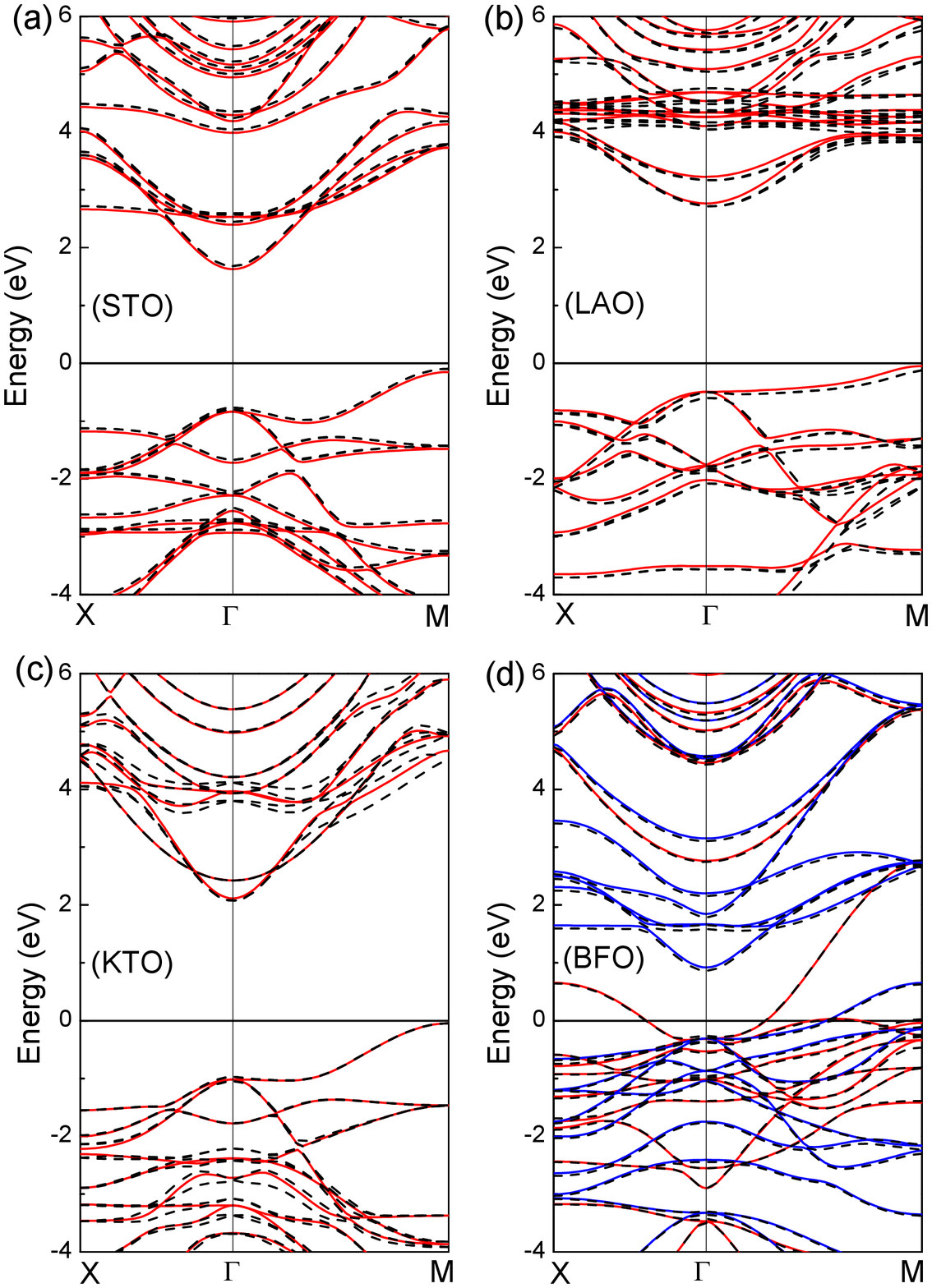}}
\caption{(Color online) The band structures of the four monolayers: STO-ML (a), LAO-ML (b), KTO-ML (c), and BFO-ML (d).  (a-c), The red solid lines represent the band structures without SOC, and the black dash lines those with SOC. (d) The red/blue solid lines represent the band structures in spin-up/down without SOC, and the black dash lines the bands with SOC.}\label{fig3}
\end{figure}

\begin{table}[!h]
\caption{The effective mass of three semiconductive monolayers at their valence-band maximum ($m_e$) and conduction-band minimum ($m_h$), with $m_0$ representing the mass of free electron.}
\begin{ruledtabular}
\begin{tabular}{cccc}
Effective mass ($m_0$) &  STO-ML  &  LAO-ML  &  KTO-ML  \\ \hline
$m_e$       &     1.05    &   1.37       &  1.22  \\
$m_h$       &     0.45    &   1.1      &  0.30
\end{tabular}
\end{ruledtabular}
\end{table}

The band structures of the four monolayers are presented in Fig. 3. For three semiconductor systems, the valence-band maximum (VBM) is located at the M point and the conduction-band minimum (CBM) at the $\Gamma$ point. When taking SOC into account, these PBE band structures change a little for the four monolayers. Actually SOC changes the HSE06 bands  only a little, too. The effective masses ($m_h$, $m_e$) of the VBM and CBM band edges for STO-ML, LAO-ML, and KTO-ML are (0.45, 1.05), (1.1, 1.37), and (0.30, 1.2) in $m_0$ (free electron mass), as summarized in Table III. It is interesting that the electron effective mass is larger than the hole effective mass for all the three semiconductor monolayers. Because BFO-ML is a ferromagnetic metal, the bands are spin-resolved and metallic. The Fermi level crosses two bands in the spin-up channnel and one band in the spin-down channel.

\begin{figure}[!htbp]
{\centering  
\includegraphics[clip, width=8cm]{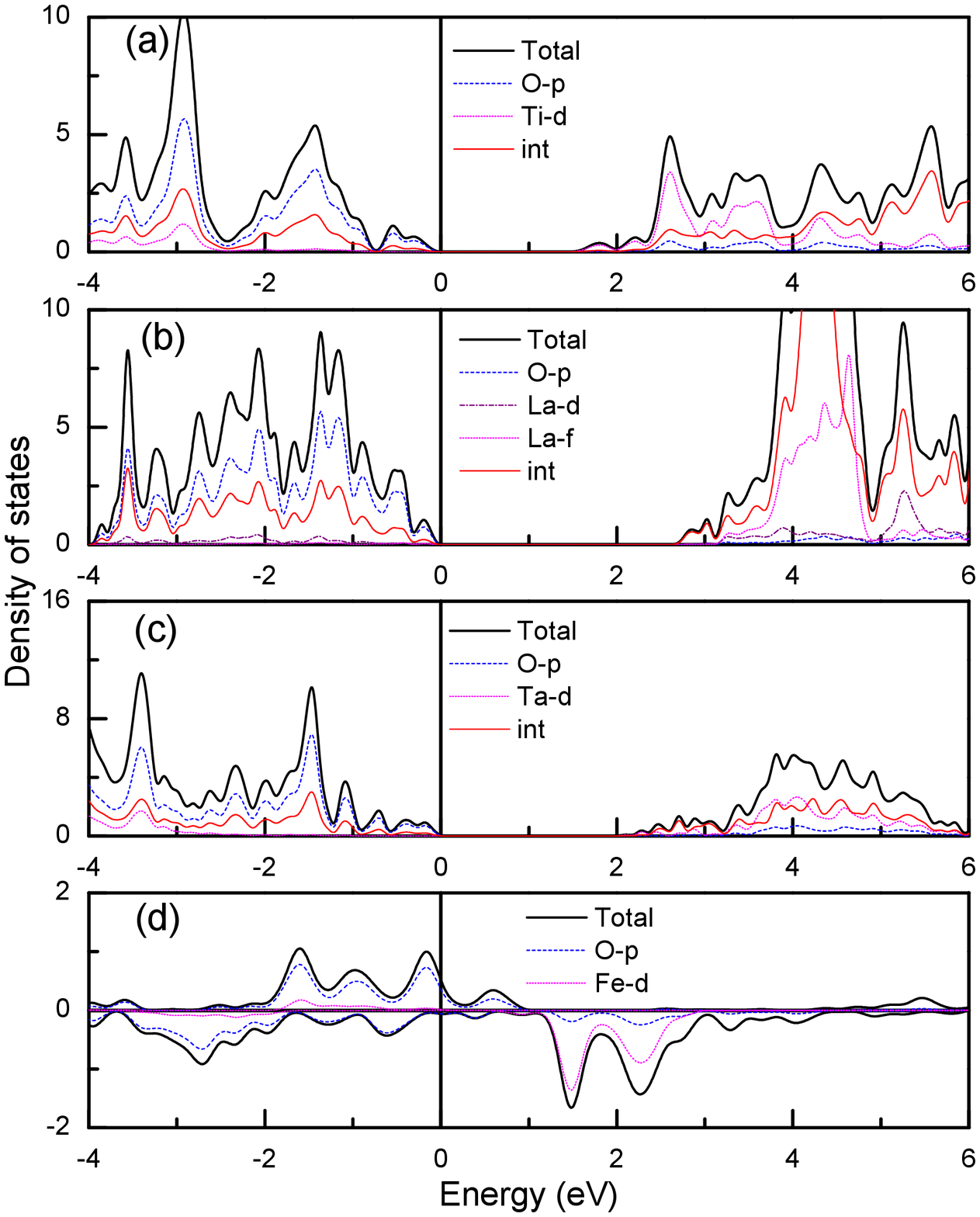}}
\caption{(Color online) The total and partial DOSs calculated with SOC and PBE of the four monolayers: STO-ML (a), LAO-ML (b), KTO-ML (c), and BFO-ML (d). For BFO-ML, the DOS is spin-polarized and the upper (lower) part represents DOS for the spin-up (spin-down) channel.}\label{fig2}
\end{figure}

Presented in Fig. 4 are the total and partial density of states (DOS) of the four monolayers. For the three semiconductor monolayers, the occupied states near the Fermi level originate from the p states of O atoms. In contrast, the empty states near the CBM have different origins. For STO-ML, they are from Ti-d states; for LAO-ML, they originate from O p states (The contribution from the interstitial region comes from O-p states, too); and for KTO-ML, they originate from Ta d and O p states, corresponding to the lowest and the next lowest conduction bands at the $\Gamma$ point. As for BFO-ML, the electronic states near the Fermi level, mainly from the spin-up channel, originate from major O p states and minor Fe d states.

\begin{figure}[!htbp]
{\centering  
\includegraphics[clip, width=8cm]{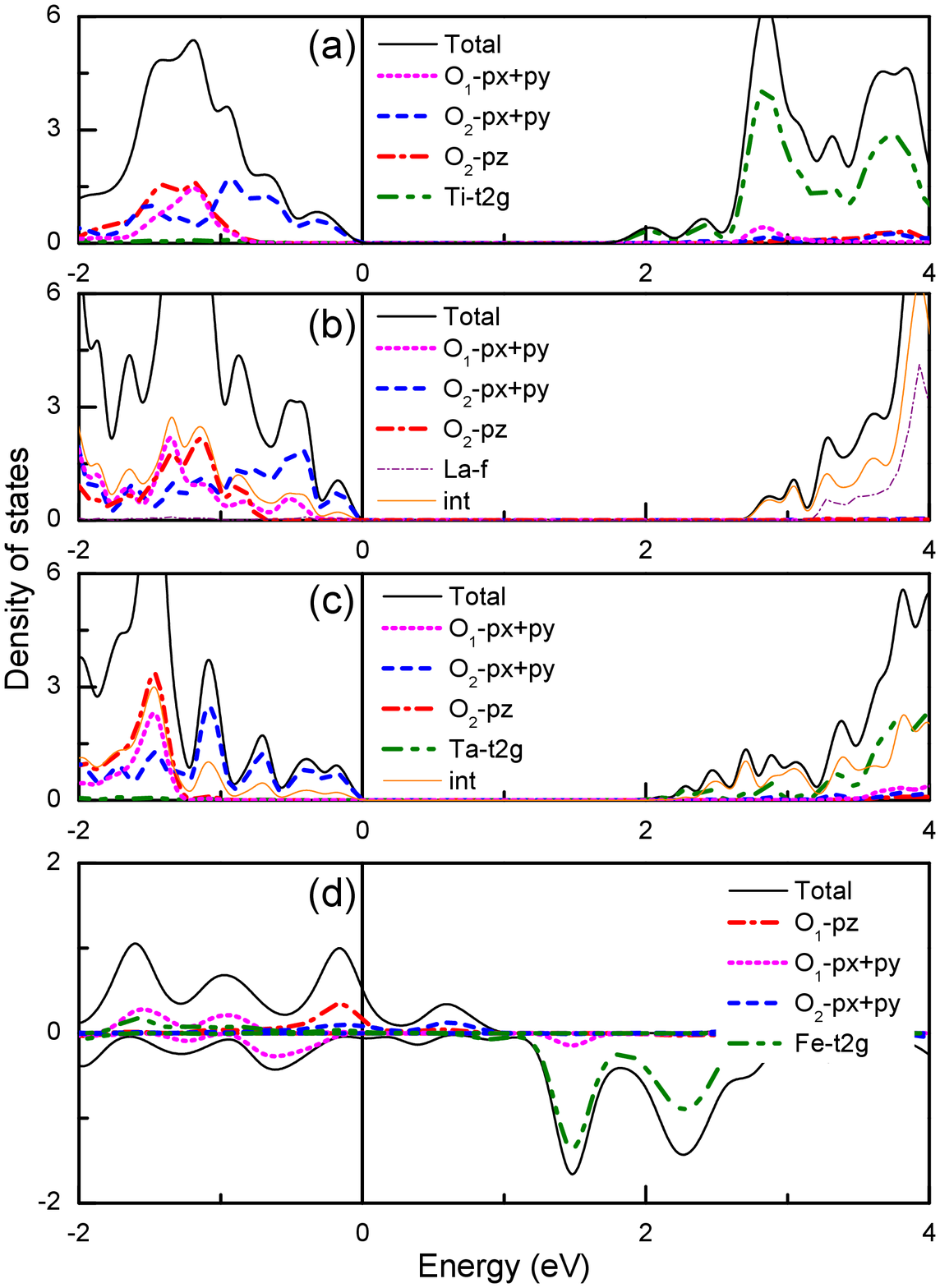}}
\caption{(Color online) The total and atom-orbital-resolved DOSs calculated with PBE and SOC of the four monolayers: STO-ML (a), LAO-ML (b), KTO-ML (c), and BFO-ML (d). For BFO-ML, the DOSs are spin-polarized and the upper (lower) part represents the DOSs for the spin-up (spin-down) channel. Here O$_1$ belongs to the A-O AL (A=Sr, La, K, Ba), and O$_2$ the B-O2 AL (B=Ti, Al, Ta, Fe).}\label{fig5}
\end{figure}

To elucidate the electronic structures, we present the orbital-resolved DOSs for the four monolayers in Fig. 5.
For STO-ML, The DOS between -0.8 eV and 0 originates from O$_2$ px+py states (Ti-O2 AL), and that between 1.8 and 2.6 eV from Ti t$_{2g}$ states, which implies that the valence and conduction band edges originate from the O and Ti atoms in the Ti-O2 AL. For LAO-ML, the DOS between -0.3 and 0 eV is contributed by O$_2$ px+py states (Al-O2 AL) only, and that between 2.7 and 3.2 eV by the interstitial region. For KTO-ML, DOS between  -1.3 and 0 eV originates from O$_2$ px+py states (Ta-O2 AL), and that between 2.1 and 3.5 eV from Ta t$_{2g}$ and the interstitial region. BFO-ML is special because DOSs below 1.1 eV originate mainly from O states, having a peak of  O$_1$ pz character near -0.2 eV, and DOSs above 1.1 eV mainly from Fe t$_{2g}$ states. The DOSs from the interstitial region should have O characters due to the large O$^{2-}$ ionic radius.

As for BFO-ML, our calculations show that a Fe atom contributes 4.4$\mu_B$ and its ferromagnetic exchange energy is equivalent to 0.044 eV per formula unit, but its magnetism is isotropic. These imply that the magnetism in BFO-ML can be described by an isotropic Heisenberg ferromagnetic model.
For STO and KTO bulks, experimental semiconductor gaps are 3.25 and 3.59 eV, and corresponding PBE (HSE) values are 1.99 (3.46) and 2.11 (3.64) eV, respectively. This comparison makes us believe that real gaps of the three semiconductor monolayers should be near 3 eV, a little smaller than the HSE06 values.


In summary, we have constructed tetragonal perovskite oxide monolayers and investigated their structural stability by using first-principles method, and then found four stable perovskite oxide monolayers: STO-ML, LAO-ML, KTO-ML, and BFO-ML. Our further study shows that first three are 2D wide-gap semiconductors, and BFO-ML is a 2D isotropic Heisenberg ferromagnet. For the three semiconductors, their valence band tops originate mainly from oxygen p orbitals, and their condunction band bottoms mainly from Ti d, O p, and Ta d orbitals, respectively. For BFO-ML, the states near the Fermi level are also mainly from oxygen p orbitals. As for issues of dangling bonds, they are removed by atomic relaxations in the three semiconductors, and there are some dangling bond effect in BFO-ML. These make a series of 2D monolayer materials. Considering various emerging phenomena in perovskite oxide heterostructures, these 2D perovskite oxide monolayer materials should be useful in future novel devices.

\begin{acknowledgments}
This work is supported by the Nature Science Foundation of China (Grant No. 11574366), by the Strategic Priority Research Program of the Chinese Academy of Sciences (Grant No.XDB07000000), and by the Department of Science and Technology of China (Grant No. 2016YFA0300701). All the numerical calculations were performed in the Milky Way \#2 Supercomputer system at the National Supercomputer Center of Guangzhou, Guangzhou, China.
\end{acknowledgments}

\end{document}